\documentclass[journal]{IEEEtran}
\usepackage{cite}
\usepackage{amsmath,amssymb,amsfonts}
\usepackage{algorithmic}
\usepackage{graphicx}
\usepackage{textcomp}
\usepackage{xcolor}
\usepackage{booktabs}
\usepackage{multirow}
\usepackage{subcaption}
\usepackage{hyperref}
\usepackage{url}
\usepackage{svg}

\begin{document}

\title{A BEV-Fusion Based Framework for Sequential Multi-Modal Beam Prediction in mmWave Systems}

\author{Jiaming Zeng, 
        Cunhua Pan, ~\IEEEmembership{Senior Member,~IEEE, }
		  Haoyang Weng,
        Ruijing Liu,
		Hong Ren, ~\IEEEmembership{Member,~IEEE, }\\
		Jiangzhou Wang ~\IEEEmembership{Fellow,~IEEE }
		\thanks{ Jiaming Zeng , Cunhua Pan, Haoyang Weng,
Ruijing Liu, ,Hong Ren are with the National Mobile Communications 
Research Laboratory, Southeast University, Nanjing 210096, China Jiangzhou 
Wang is with the National Mobile Communications Research Laboratory, 
Southeast University, Nanjing 210096,China, and also with the Pervasive 
Communication Research Center, Purple Mountain Laboratories,(E-mail:
220250992@seu.edu.cn; cpan@seu.edu.cn; 220241245@seu.edu.cn;
230258158@seu.edu.cn;hren@seu.edu.cn: j.z.wang@seu.edu.cn)
.
        }
        }
\maketitle

\begin{abstract}
Beam prediction is critical for reducing beam-training overhead in millimeter-wave (mmWave) systems, especially in high-mobility vehicular scenarios. This paper presents a BEV-Fusion based framework that unifies camera, LiDAR, radar, and GPS modalities in a shared bird's-eye-view (BEV) representation for spatially consistent multi-modal fusion. Unlike prior approaches that fuse globally pooled one-dimensional features, the proposed method performs fusion in BEV space to preserve cross-modal geometric structure and visual semantic density. A learned camera-to-BEV module based on cross-attention is adopted to generate BEV-aligned visual features without relying on precise camera calibration, and a temporal transformer is used to aggregate five-step sequential observations for motion-aware beam prediction. Experiments on the DeepSense 6G benchmark show that BEV-Fusion achieves approximately 87\% distance-based accuracy (DBA) on scenarios 32, 33 and 34, outperforming the TransFuser baseline. These results indicate that BEV-space fusion provides an effective spatial abstraction for sensing-assisted beam prediction.
\end{abstract}

\begin{IEEEkeywords}
Millimeter wave communications, beam prediction, multi-modal fusion, bird's-eye view representation, deep learning,
\end{IEEEkeywords}

\section{Introduction}

The rapid evolution of millimeter-wave (mmWave) and terahertz (THz) communications for 5G-Advanced and emerging 6G systems has created strict requirements for high-throughput and low-latency wireless links \cite{rappaport2019wireless}. To compensate for severe propagation loss at high frequencies, these systems rely on large-scale antenna arrays and highly directional beams \cite{heath2016overview}. However, identifying the best beam from a predefined codebook through beam training incurs substantial overhead that increases with codebook size and user mobility \cite{alkhateeb2014channel}. 
In practical deployments, beam training overhead can consume 
20-40\% of available time-frequency resources for codebooks with 64-256 beams 
\cite{alkhateeb2018deep}. For instance, in a 28 GHz system with 100 MHz 
bandwidth and 64-beam codebook, exhaustive beam sweeping requires approximately 
6.4 ms (assuming 100 µs per beam measurement), which reduces effective data 
transmission time by 15-30\% depending on coherence time constraints. In 
vehicular scenarios, where users move rapidly in dynamic environments (typical 
speeds 60-120 km/h), beam coherence time drops to 100-500 ms \cite{zhang2021reinforcement}, 
necessitating frequent beam updates that further exacerbate this overhead. 
This overhead becomes a key bottleneck for reliable real-time mmWave/THz 
connectivity, motivating the need for efficient beam prediction techniques.

Sensing-assisted beam prediction has therefore emerged as an attractive alternative, where environmental side information is used to infer the optimal beam without exhaustive beam sweeping \cite{alkhateeb2018deep,zhang2021reinforcement}. A base station (BS) equipped with camera, LiDAR, radar, and GPS sensors can capture complementary observations of user position, motion, and surrounding geometry. 

\textbf{Single-Modal Approaches}: Vision-based methods exploit the relationship 
between scene content and beam direction, and convolutional neural networks 
(CNNs) have reported beam index prediction accuracies above 75\% under 
controlled settings \cite{alrabeiah2020millimeter,zhang2023vision}. LiDAR-based 
approaches utilize 3D point clouds for geometric characterization \cite{jiang2022lidar}, 
achieving 65-70\% accuracy by leveraging explicit depth information. Radar-based 
approaches provide robust motion cues under adverse weather conditions 
\cite{demirhan2022radar}, with particular strength in detecting user velocity 
(Doppler information). Position-based approaches use GPS or relative coordinates 
as lightweight channel surrogates with competitive performance \cite{morais2022position}, 
reaching 70-75\% accuracy in line-of-sight scenarios.

\textbf{Multi-Modal Fusion}: Recent multi-modal methods that fuse visual and 
positional cues have further improved performance on the DeepSense 6G benchmark, 
with reported distance-based accuracy (DBA) above 78\% \cite{tian2023multimodal}. 
Charan et al. \cite{charan2022vision} combined vision and position features 
via concatenation in 1D latent space. Prakash et al. \cite{prakash2021multifusion} 
introduced TransFuser, which applies transformer-based fusion after separate 
modality-specific encoding, achieving 77.64\% DBA on scenarios 32-34. However, 
these approaches share a common limitation: they perform fusion in compressed 
1D feature spaces after aggressive spatial pooling.

Despite this progress, two major limitations remain in existing multi-modal beam prediction frameworks. First, most methods perform feature fusion in compressed one-dimensional (1D) latent spaces after global pooling, which discards spatial structure in each modality before cross-modal interaction. As a result, camera features are reduced to global descriptors, LiDAR structure is overly compressed, and explicit geometric correspondence across modalities is weakened. This limitation is similar to early point-level fusion in autonomous perception \cite{vora2020pointpainting,wang2021pointaugmenting}, where geometric detail and semantic density are not fully preserved. Second, current prediction performance is still insufficient for robust deployment in safety-critical vehicular scenarios; specifically, existing methods have not yet achieved consistently high DBA under complex mobility and sensing conditions.

\begin{figure}
    \centering
    \includegraphics[width=1\linewidth]{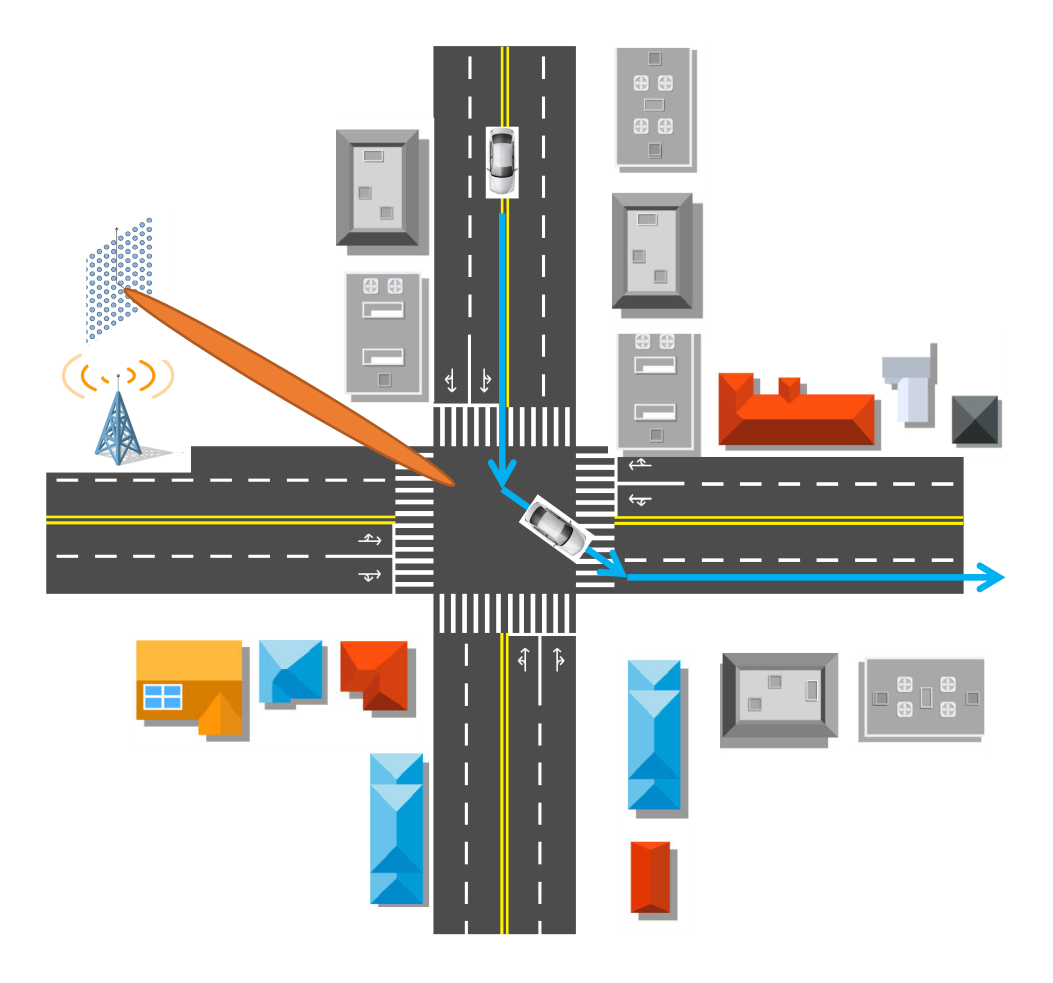}
    \caption{Mobile mmWave communication system}
    \label{fig:intro}
\end{figure}
Motivated by the success of Bird's-Eye View (BEV) representation in autonomous driving perception \cite{liu2023bevfusion,li2022bevformer}, we propose a BEV-Fusion framework for multi-modal beam prediction. The key idea is to map camera, LiDAR, radar, and GPS observations into a unified BEV coordinate system and perform fusion directly in that shared spatial domain. Compared with 1D feature fusion, BEV-space fusion preserves camera semantic density and LiDAR/radar geometric structure, enabling spatially grounded cross-modal interaction. Prior perception studies have shown that BEV representations provide strong geometric consistency and rich semantics for multi-sensor reasoning tasks .

The main contributions of this work are summarized as follows:

\begin{itemize}
    \item \textbf{Unified BEV Framework for Beam Prediction}: We propose a BEV-based multi-modal framework for mmWave beam prediction that projects camera, LiDAR, radar, and GPS observations into a shared spatial plane for joint fusion. Compared with latent-space fusion methods \cite{tian2023multimodal}, the framework preserves explicit spatial correspondence across modalities throughout the pipeline.

    \item \textbf{Learned Camera-to-BEV Transformation}: We adopt a cross-attention mapping with learnable BEV queries attending to 2D camera features extracted by ResNet-34 \cite{he2016deep}. This design avoids dependence on precise camera calibration while producing dense BEV-aligned visual features, unlike geometry-driven methods that require known intrinsics/extrinsics \cite{philion2020lift}.

     \item \textbf{Dual-Pathway GPS-to-BEV Encoding}: To address the extreme sparsity of GPS-derived spatial 
    representations, we propose a dual-pathway encoding strategy that combines a spatial BEV mask with a parallel global MLP pathway. The spatial pathway preserves explicit grid-level localization for direct fusion with other BEV modalities, while the global pathway encodes full-precision positional information via a gated residual injection, compensating for the information loss caused by binary mask sparsity. This 
    design is more robust than single-pathway GPS encoding used in prior works \cite{morais2022position}.

    \item \textbf{Temporal Sequential Modeling}: Unlike prior methods that apply temporal modeling directly 
    on raw modality features or 1D latent vectors \cite{charan2022vision, tian2023multimodal}, we propose to 
    perform temporal fusion after spatial BEV fusion, such that the transformer operates on geometrically consistent multi-modal representations rather than unimodal or unstructured sequences. This design enables the temporal module to capture motion-dependent beam evolution with full spatial context preserved across all modalities, supporting trajectory-level reasoning that is not achievable by per-modality sequential models.

   \item \textbf{Performance Gains on DeepSense 6G}:Experiments on the DeepSense 6G benchmark \cite{charan2023deepsense} demonstrate that our method achieves \textbf{86.52\%} overall DBA 
on scenarios 32--34, outperforming the TransFuser baseline by \textbf{8.88} percentage points.
\end{itemize}

The remainder of this paper is organized as follows. Section~\ref{sec:related work} reviews 
related work on sensing-assisted beam prediction and BEV representations. Section~\ref{sec:system_model} introduces the system model and problem formulation. Section~\ref{sec:method} details the proposed BEV-Fusion architecture. Section~\ref{sec:experiments} describes the experimental setup. Section~\ref{sec:results} reports the performance evaluation and ablation results. Section~\ref{sec:conclusion} concludes the paper.

\section{Related Work}
\label{sec:related work}
\subsection{Multi-Modal Fusion for Beam Prediction}

The different sensing modalities for beam prediction 
has evolved through three distinct architectural paradigms. Early approaches 
employed independent feature extraction followed by late fusion 
\cite{charan2022vision}, wherein camera, LiDAR, and GPS observations were 
encoded separately through modality-specific networks (e.g., ResNet for 
vision, PointNet for LiDAR, MLP for position), then concatenated after 
global average pooling. While computationally efficient, this strategy 
precludes cross-modal interaction prior to spatial downsampling, thereby 
limiting the model's capacity to exploit geometric correspondences across 
modalities.

Recent attention-based architectures \cite{tian2023multimodal} 
have demonstrated improved performance through transformer-based fusion 
mechanisms. The TransFuser framework \cite{prakash2021multifusion}, adopted 
as the DeepSense 6G baseline, applies multi-head self-attention to 
concatenated modality tokens, achieving 77.64\% distance-based accuracy (DBA). 
Despite enabling learned cross-modal dependencies, these methods operate on 
globally-pooled one-dimensional representations—spatial correspondence is not 
explicitly preserved during fusion. Specifically, the model cannot directly 
reason about the alignment between camera pixel coordinates and LiDAR/radar 
spatial positions, as all modalities are projected to unstructured feature 
vectors prior to interaction.

The proposed BEV-Fusion based framework departs from this latent-space fusion 
paradigm by performing cross-modal integration in a unified two-dimensional 
Bird's-Eye View coordinate system. This architectural choice preserves 
explicit spatial structure throughout the fusion pipeline, enabling 
geometrically grounded multi-modal reasoning.

\subsection{Bird's-Eye View Representations in Multi-Sensor Perception}

BEV representations have emerged as a dominant paradigm for multi-sensor 
fusion in autonomous driving applications \cite{liu2023bevfusion,li2022bevformer,philion2020lift}, 
where unified spatial representations facilitate joint processing of camera, 
LiDAR, and radar observations for 3D object detection and motion forecasting.

Two key technical innovations inform our design. First, learned camera-to-BEV 
transformations eliminate dependence on precise camera 
calibration through cross-attention mechanisms: a set of learnable BEV query 
embeddings attend to perspective-view camera features, implicitly learning 
the geometric transformation from supervision alone. This calibration-free 
property is particularly advantageous for mmWave base station deployments 
where camera extrinsic parameters may be unavailable or subject to change. 
Second, unified BEV fusion has been shown to outperform 
feature-level or decision-level fusion by 2-4\% in mean average precision on 
standard benchmarks, attributed to explicit preservation of geometric structure 
during cross-modal interaction.

However, direct transfer of these techniques to beam prediction presents 
three fundamental challenges: (1) \textit{Supervision granularity}—autonomous 
driving employs dense spatial annotations (bounding boxes, semantic masks), 
whereas beam prediction provides only class-level supervision (beam indices) 
without explicit spatial guidance; (2) \textit{Sensor heterogeneity}—mmWave 
systems incorporate radar Doppler velocity and GPS positioning, which lack 
established BEV encoding strategies in the driving literature; (3) 
\textit{Temporal modeling}—beam prediction necessitates sequential aggregation 
across multiple timesteps to capture user trajectory, whereas existing BEV 
methods predominantly target single-frame perception tasks.


\subsection{Temporal Modeling Strategies for Sequential Prediction}

Sequential beam prediction requires effective aggregation of multi-timestep 
observations to capture vehicular motion patterns and channel evolution 
dynamics. Existing approaches can be categorized into two classes:

\textit{Per-Modality Temporal Fusion}: Charan et al. \cite{charan2022vision} 
employed recurrent neural networks (LSTMs) to process sequential camera frames, 
subsequently fusing temporal features with static GPS coordinates. While this 
design captures intra-modality temporal dependencies, it cannot model 
cross-modal evolution patterns—for instance, the correspondence between 
visual trajectory and radar velocity profiles over time.

\textit{Post-Fusion Temporal Aggregation}: The TransFuser baseline 
\cite{prakash2021multifusion} applies transformer-based temporal modeling 
to fused feature sequences. However, fusion at each timestep is performed 
via global pooling to one-dimensional vectors, such that spatial structure 
is discarded prior to temporal aggregation. Consequently, the temporal 
transformer operates on abstract feature embeddings rather than 
spatially-grounded representations.

In contrast, BEV-Fusion performs temporal modeling subsequent to
spatial BEV fusion (Section IV-E). At each timestep $t \in \{1,\ldots,5\}$, 
multi-modal observations are first transformed to a unified BEV representation 
$\mathbf{F}_{\text{bev},t} \in \mathbb{R}^{256 \times 128 \times 128}$ 
preserving geometric alignment across camera, LiDAR, radar, and GPS. The 
temporal transformer then aggregates the sequence $\{\mathbf{F}_{\text{bev},t}\}_{t=1}^5$ 
after spatial global pooling, enabling trajectory-level reasoning where 
cross-modal spatial correspondence is preserved throughout sequential 
processing. This architectural choice represents a departure from prior 
temporal fusion strategies, which operate on either unimodal sequences or 
spatially-collapsed multi-modal representations.

\subsection{Summary}

Table~\ref{tab:related_positioning} provides a comparative summary of 
BEV-Fusion relative to representative prior work across key architectural 
dimensions.

\begin{table}[t]
\caption{Architectural Comparison of Multi-Modal Beam Prediction Methods}
\centering
\small
\begin{tabular}{p{2.2cm}ccc}
\toprule
\textbf{Method} & \textbf{Fusion} & \textbf{Spatial} & \textbf{Temporal} \\
 & \textbf{Domain} & \textbf{Structure} & \textbf{Stage} \\
\midrule
Charan  & 1D Concat & Not preserved & Pre-fusion \\
TransFuser & 1D Latent & Not preserved & Post-fusion \\
BEVFormer$^\dagger$  & 2D BEV & Preserved & Single-frame \\
\textbf{BEV-Fusion(ours)} & \textbf{2D BEV} & \textbf{Preserved} & \textbf{Post-spatial} \\
\bottomrule
\multicolumn{4}{l}{\scriptsize $^\dagger$Autonomous driving baseline, not designed for beam prediction} \\
\end{tabular}
\label{tab:related_positioning}
\end{table}

To the best of our knowledge, the proposed framework represents the first 
application of unified BEV spatial fusion to mmWave beam prediction, combining 
calibration-free camera-to-BEV transformation with temporal modeling over 
geometrically aligned multi-modal observation sequences.
\section{System Model and Problem Formulation}
\label{sec:system_model}






\subsection{Multi-Modal Sensing Model}

To avoid exhaustive beam training, the BS is equipped with a multi-modal sensing suite comprising four complementary sensor types:

\textbf{Camera}: An RGB camera captures front-view images $\mathbf{I}_t \in \mathbb{R}^{H \times W \times 3}$ at each time step $t$, where $H$ and $W$ denote image height and width respectively. The camera provides rich semantic information about the visual scene, including user appearance and surrounding environment structure .

\textbf{LiDAR}: A LiDAR sensor generates 3D point clouds of the environment. Following standard preprocessing \cite{qi2017pointnet}, each point cloud is projected onto a Bird's-Eye View (BEV) grid, yielding a 2D spatial map $\mathbf{L}_t \in \mathbb{R}^{H_l \times W_l \times C_l}$ encoding height, intensity, and density in each grid cell. The BEV projection preserves geometric structure while reducing computational complexity.

\textbf{Radar}: A mmWave radar provides Range-Angle (RA) and Range-Velocity (RV) maps via 2D Fourier transform processing \cite{demirhan2022radar}, concatenated as $\mathbf{R}_t \in \mathbb{R}^{H_r \times W_r \times 2}$. The velocity channel provides motion information unavailable to camera and LiDAR, particularly valuable under challenging lighting conditions.

\textbf{GPS}: A GPS receiver at the UE provides 2D position measurements $\mathbf{g}_t = (\Delta x_t, \Delta y_t)$, representing the relative Cartesian displacement between UE and BS \cite{morais2022position}. Due to practical collection constraints in the DeepSense 6G dataset, GPS measurements are available only for the first two time instances in each sequence.

Based on the above configuration,the BS collects synchronized observations across $T = 5$ sequential time steps, forming a temporal multi-modal observation sequence
\begin{equation}
\mathcal{S} = \{(\mathbf{I}_t, \mathbf{L}_t, \mathbf{R}_t)\}_{t=1}^{T} \cup \{(\mathbf{g}_t)\}_{t=1}^{2}.
\end{equation}

\subsection{Beam Prediction Problem Formulation}

Given the multi-modal observation sequence $\mathcal{S}$, the beam prediction task is formulated as a multi-class classification problem: learn a prediction function $f_\Theta(\mathcal{S})$ that maps the multi-modal sensory observations to a probability distribution over the $M$ candidate beams,and can be formulated as:
\begin{equation}
\hat{P} = f_\Theta(\mathcal{S}) \in \mathbb{R}^M, \quad \sum_{m=1}^M \hat{P}_m = 1.
\end{equation}

The predicted optimal beam index can be obtained by
\begin{equation}
\hat{m}^* = \arg\max_{m \in \{1,...,M\}} \hat{P}_m.
\end{equation}

The model parameters $\Theta$ are learned from a labeled training dataset $\mathcal{D} = \{(\mathcal{S}^{(i)}, m^{*(i)})\}_{i=1}^{N}$ .

\textbf{Evaluation Metric}: Following the DeepSense 6G challenge protocol\cite{charan2023deepsense}, model performance is evaluated using the Distance-Based Accuracy (DBA) score, which accounts for the structured nature of beam codebooks by rewarding predictions close to the ground truth,which can be formulated as
\begin{equation}
\text{DBA} = \frac{1}{K} \sum_{k=1}^{K} Y_k,
\end{equation}
where
\begin{equation}
Y_k = 1 - \frac{1}{N} \sum_{n=1}^{N} \min_{1 \leq j \leq k} \min\left(\frac{|\hat{m}^{(n)}_j - m^{*(n)}|}{\Delta}, 1\right),
\end{equation}
with $K$ stands for top 3 predicted beams, $\Delta = 5$ as the distance normalization factor, and $\hat{m}^{(n)}_j$ denotes the $j$-th most likely predicted beam for sample $n$. The DBA score ranges from 0 to 1, where higher values indicate better performance.

\textbf{Key Challenge}: The fundamental challenge addressed in this work is achieving high beam prediction accuracy in well-represented deployment scenarios through effective multi-modal fusion. In the DeepSense 6G dataset, scenarios 32, 33, and 34 provide abundant labeled training data ($\sim$11,000 samples), 
We design a \textbf{unified Bird's-Eye View (BEV) spatial representation},which can serve as a more effective fusion domain, since the top-down BEV coordinate system preserves both geometric structure (from LiDAR/radar) and semantic density (from camera), enabling richer cross-modal interactions.

\section{Proposed BEV-Fusion Based Framework for Beam Prediction}
\label{sec:method}

\begin{figure*}
    \centering
    \includegraphics[width=1\linewidth]{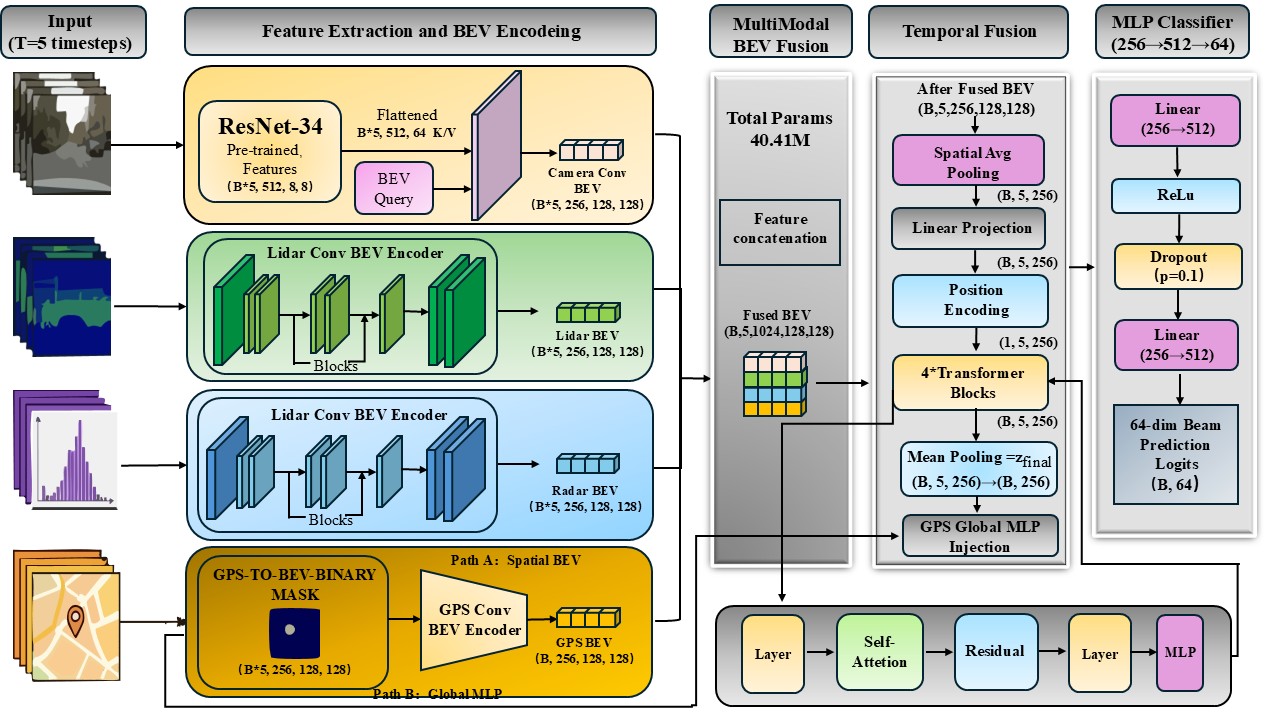}
    \caption{Modal Architecture}
    \label{fig:architecture}
\end{figure*}

This section presents the proposed BEV-Fusion based framework, which consists of five main components: (1) multi-modal data preprocessing; (2) modality-specific BEV encoders; (3) unified BEV fusion; (4) temporal transformer; (5) beam classification head. The overall architecture is shown in Fig.~\ref{fig:architecture} at the top of the next page.

\subsection{Framework Overview}

The BEV-Fusion based framework transforms multi-modal sensory data into a unified Bird's-Eye View representation before performing fusion.
 The key design principle is to preserve spatial structure throughout the feature extraction and fusion pipeline, rather than collapsing features to 1D vectors prematurely as done in prior works \cite{prakash2021multifusion}. The architecture processes each modality through specialized encoders that output features in the same BEV coordinate system with dimensions $(B, C_{bev}, H_{bev}, W_{bev})$, where $B$ is batch size, $C_{bev}=256$ is the feature dimension, and $(H_{bev}, W_{bev}) = (128, 128)$ defines the BEV spatial grid covering a $[-50m, 50m] \times [-50m, 50m]$ region around the BS.

\subsection{Multi-Modal Data Preprocessing}

\subsubsection{Camera Data Normalization}

Camera images $\mathbf{I}_t \in \mathbb{R}^{H \times W \times 3}$ are 
first resized to $256 \times 256$ pixels via bilinear interpolation, 
then normalized using ImageNet statistics \cite{he2016deep,simonyan2015very}
\begin{equation}
\mathbf{I}_t^{\text{norm}} = \frac{\mathbf{I}_t / 255 - \boldsymbol{\mu}}
{\boldsymbol{\sigma}},
\end{equation}

where the channel-wise mean and standard deviation are
\begin{equation}
\begin{aligned}
\boldsymbol{\mu} = [0.485,\, 0.456,\, 0.406]^T ,\\
\boldsymbol{\sigma} = [0.229,\, 0.224,\, 0.225]^T.
\end{aligned}
\end{equation}

This normalization ensures compatibility with ResNet-34 pre-trained 
weights and stabilizes gradient flow during training.

\subsubsection{LiDAR BEV Projection}

Raw 3D LiDAR point clouds $\mathcal{P}_t = \{(\mathbf{x}_i, \mathbf{y}_i, 
\mathbf{z}_i, I_i)\}_{i=1}^{N_p}$ are projected onto a 2D Bird's-Eye View 
(BEV) grid following established autonomous driving practices 
\cite{zhou2018voxelnet}. The BEV grid spans 
$[-50\text{m}, +50\text{m}]^2$ with resolution $128 \times 128$, yielding 
a spatial resolution of approximately $0.78\text{m}$ per grid cell.

For each grid cell $(u,v)$, we encode three complementary features by 
aggregating points falling within the cell
\begin{equation}
\begin{aligned}
h_{u,v} &= \max_{(x_i, y_i) \in \text{cell}(u,v)} z_i
\quad &\text{(max height)}, \\
e_{u,v} &= \max_{(x_i, y_i) \in \text{cell}(u,v)} I_i
\quad &\text{(max intensity)}, \\
d_{u,v} &= \sum_{i} 1\left[(x_i, y_i) \in \text{cell}(u,v)\right]
\quad &\text{(point density)}
\end{aligned}
\end{equation}
where $\text{cell}(u,v)$ denotes the spatial region corresponding to grid 
position $(u,v)$. These three channels are stacked to form the LiDAR BEV 
representation
\begin{equation}
\mathbf{L}_t = [h_{u,v}, e_{u,v}, d_{u,v}] \in \mathbb{R}^{H_l \times W_l \times 3}.
\end{equation}

However, in our implementation following the DeepSense 6G dataset format, 
we use a single-channel height-encoded BEV projection $\mathbf{L}_t \in 
\mathbb{R}^{1 \times H_l \times W_l}$ for computational efficiency. This 
is then bilinearly interpolated to the canonical BEV grid size
\begin{equation}
\tilde{\mathbf{L}}_t = \text{Interp}_{\text{bilinear}}(\mathbf{L}_t, 
128 \times 128) \in \mathbb{R}^{B \times 1 \times 128 \times 128}.
\end{equation}

\subsubsection{Radar FFT Processing}

Raw radar signals undergo 2D Fast Fourier Transform (FFT) processing 
along the fast-time and slow-time dimensions to extract Range-Angle (RA) 
and Range-Velocity (RV) maps \cite{demirhan2022radar}. Let $\mathbf{S}_t 
\in \mathbb{C}^{N_r \times N_c}$ denote the complex-valued radar cube 
after pulse compression, where $N_r$ and $N_c$ represent the number of 
range bins and chirps respectively. The RA map is obtained via FFT along 
the spatial dimension
\begin{equation}
\text{RA}_t[r,\theta] = \left|\text{FFT}_{\text{spatial}}
(\mathbf{S}_t[r,:])\right|^2.
\end{equation}

and the RV map via FFT along the temporal dimension
\begin{equation}
\text{RV}_t[r,v] = \left|\text{FFT}_{\text{Doppler}}
(\mathbf{S}_t[r,:])\right|^2,
\end{equation}

where $|\cdot|$ denotes magnitude. These two maps capture complementary 
information: RA provides spatial localization (azimuth angle and range), 
while RV encodes motion characteristics (radial velocity). The two maps 
are concatenated along the channel dimension
\begin{equation}
\mathbf{R}_t = [\text{RA}_t \;\|\; \text{RV}_t] \in \mathbb{R}^{H_r \times 
W_r \times 2}.
\end{equation}

and bilinearly interpolated to match the BEV grid size
\begin{equation}
\tilde{\mathbf{R}}_t = \text{Interp}_{\text{bilinear}}(\mathbf{R}_t, 
128 \times 128) \in \mathbb{R}^{B \times 2 \times 128 \times 128}.
\end{equation}

\subsubsection{GPS Coordinate Calibration}
GPS measurements provide 2D relative position $\mathbf{g} = (\Delta x, 
\Delta y)$ in Cartesian coordinates. Following \cite{morais2022position}, 
we apply scenario-specific angular calibration to align the GPS coordinate 
frame with the camera field of view. Let $\theta_{\text{offset}}$ denote 
the calibration angle for a given scenario; the calibrated GPS coordinates 
are
\begin{equation}
\begin{bmatrix} \Delta x' \\ \Delta y' \end{bmatrix} = 
\begin{bmatrix} 
\cos\theta_{\text{offset}} & -\sin\theta_{\text{offset}} \\ 
\sin\theta_{\text{offset}} & \phantom{-}\cos\theta_{\text{offset}} 
\end{bmatrix}
\begin{bmatrix} \Delta x \\ \Delta y \end{bmatrix}
\end{equation}

The calibrated coordinates $(\Delta x', \Delta y')$ are then used for 
subsequent BEV mask generation (Section III-C-3) and global embedding.
\subsection{Modality-Specific BEV Encoders}

\subsubsection{Camera-to-BEV Transformation via Cross-Attention}

Unlike autonomous driving applications where camera intrinsic and extrinsic parameters are precisely calibrated \cite{philion2020lift}, wireless BS deployments may lack such calibration data. We adopt a learned camera-to-BEV transformation using cross-attention mechanisms inspired by BEVFormer \cite{li2022bevformer}.

\textbf{Step 1: Camera Feature Extraction.}
A ResNet-34 backbone \cite{he2016deep} pretrained on ImageNet 
\cite{simonyan2015very} extracts hierarchical visual features from the 
normalized camera image $\mathbf{I}_t^{\text{norm}}$
\begin{equation}
\mathbf{F}_t^{cam} = \text{ResNet34}(\mathbf{I}_t^{\text{norm}}) 
\in \mathbb{R}^{B \times 512 \times 8 \times 8}.
\end{equation}

The output feature map has spatial resolution $8 \times 8$ (corresponding 
to $32\times$ downsampling from the input $256 \times 256$ image) with 
512 channels. These features are then flattened along the spatial dimension
\begin{equation}
\mathbf{f}_t^{cam} = \text{Flatten}_{spatial}(\mathbf{F}_t^{cam}) 
\in \mathbb{R}^{B \times 64 \times 512},
\end{equation}

where $64 = 8 \times 8$ represents the number of spatial positions that 
will serve as keys and values in the subsequent cross-attention.

\textbf{Step 2: BEV Query Generation.}
We define a set of learnable BEV query embeddings $\mathbf{E}_{bev} \in 
\mathbb{R}^{H_{bev} \times W_{bev} \times C_{bev}}$ and learnable 2D 
positional encodings $\mathbf{P}_{bev} \in \mathbb{R}^{H_{bev} \times W_{bev} 
\times C_{bev}}$ , where 
$(H_{bev}, W_{bev}) = (128, 128)$ and $C_{bev} = 256$. The BEV queries 
are constructed as
\begin{equation}
\mathbf{Q}_{bev} = \mathbf{E}_{bev} + \mathbf{P}_{bev} \in \mathbb{R}^{H_{bev} 
\times W_{bev} \times C_{bev}}.
\end{equation}

After flattening the spatial dimensions

\begin{equation}
\begin{split}
\mathbf{Q}_{bev}^{flat} &= \text{Flatten}_{spatial}(\mathbf{Q}_{bev}) \\
&\in \mathbb{R}^{B \times (H_{bev} \cdot W_{bev}) \times C_{bev}} \\
&= \mathbb{R}^{B \times 16384 \times 256}.
\end{split}
\end{equation}

Each of the $16384 = 128 \times 128$ query vectors corresponds to a 
specific BEV grid cell and will attend to the camera features to aggregate 
relevant visual information.

\textbf{Step 3: Cross-Attention Mechanism.}
The camera features are projected to match the query dimension via a 
linear transformation
\begin{equation}
\begin{aligned}
\mathbf{K}_t= \mathbf{W}_K \mathbf{f}_t^{cam}, \in \mathbb{R}^{B \times 64 \times 256} \\
\mathbf{V}_t =\mathbf{W}_V \mathbf{f}_t^{cam} \in \mathbb{R}^{B \times 64 \times 256},
\end{aligned}
\end{equation}

where $\mathbf{W}_K, \mathbf{W}_V \in \mathbb{R}^{256 \times 512}$ are 
learnable projection matrices. Multi-head cross-attention is then applied
\begin{equation}
\text{Attn}(\mathbf{Q}, \mathbf{K}, \mathbf{V}) = \text{softmax}\left(
\frac{\mathbf{Q}\mathbf{K}^T}{\sqrt{d_k}}\right) \mathbf{V}.
\end{equation}

where $d_k = C_{bev} / N_h = 256 / 4 = 64$ is the dimension per attention 
head for $N_h = 4$ heads. The output of each attention head is computed as
\begin{equation}
\text{head}_k = \text{softmax}\!\left(\frac{Q_{bev} K_t^\top}{\sqrt{d_k}}\right)V_t, 
\quad k=1,\dots,N_h.
\end{equation}

    The outputs of all heads are concatenated and projected via a 
learnable matrix $W^O \in \mathbb{R}^{C_{bev}\times C_{bev}}$:
\begin{equation}
Z = \text{Concat}(\text{head}_1, \dots, \text{head}_{N_h})\,W^O 
\in \mathbb{R}^{B \times (H_{bev}\cdot W_{bev}) \times C_{bev}}.
\end{equation}

A residual connection is then applied by adding the original 
BEV query $Q_{bev}^{flat}$
\begin{equation}
\tilde{Q}_{bev} = Q_{bev}^{flat} + Z.
\end{equation}

Finally, layer normalization is applied to produce the 
updated BEV features
\begin{equation}
Q_{bev}' = \text{LN}\!\left(\tilde{Q}_{bev}\right),
\end{equation}

where $\mathbf{W}_O \in \mathbb{R}^{256 \times 256}$ is the output 
projection matrix. We apply 3 stacked cross-attention layers with identical 
architecture, and the final output is reshaped to recover the BEV spatial 
structure

\begin{equation}
\begin{split}
\mathbf{F}_{bev,t}^{cam} &= \text{Reshape}(\mathbf{Q}_{bev}^{(3)}, [B, C_{bev}, H_{bev}, W_{bev}]) \\
&\quad \in \mathbb{R}^{B \times 256 \times 128 \times 128}.
\end{split}
\end{equation}

This learned transformation avoids dependence on camera calibration while 
producing dense BEV-aligned visual features, contrasting with geometry-driven 
methods that require known intrinsics and extrinsics \cite{philion2020lift}.

\subsubsection{LiDAR and Radar BEV Encoding}



Both LiDAR and radar data are already in 2D spatial formats aligned with 
the ground plane \cite{demirhan2022radar}. We apply 
structurally identical lightweight convolutional encoders to transform them 
into the unified BEV feature space
\begin{equation}
\mathbf{F}_{bev,t}^{lid} = \text{Enc}_{lid}(\tilde{\mathbf{L}}_t) 
\in \mathbb{R}^{B \times 256 \times 128 \times 128},
\end{equation}
\begin{equation}
\mathbf{F}_{bev,t}^{rad} = \text{Enc}_{rad}(\tilde{\mathbf{R}}_t) 
\in \mathbb{R}^{B \times 256 \times 128 \times 128},
\end{equation}

where $\tilde{\mathbf{L}}_t \in \mathbb{R}^{B \times 1 \times 128 \times 128}$ 
is the bilinearly interpolated LiDAR BEV projection, and $\tilde{\mathbf{R}}_t 
\in \mathbb{R}^{B \times 2 \times 128 \times 128}$ is the interpolated radar 
Range-Angle and Range-Velocity map concatenation.

Each encoder consists of three convolutional blocks with batch normalization and ReLU activation, progressively increasing channel 
dimensions while maintaining spatial resolution
\begin{equation}
\label{eq:lidar_encoder}
\begin{aligned}
\mathbf{h}_1 &= \text{ReLU}(\text{BN}(\text{Conv}_{3 \times 3}^{C_{in} \to 64}
(\tilde{\mathbf{X}}_t))) \in \mathbb{R}^{B \times 64 \times 128 \times 128} \\
\mathbf{h}_2 &= \text{ReLU}(\text{BN}(\text{Conv}_{3 \times 3}^{64 \to 128}
(\mathbf{h}_1))) \in \mathbb{R}^{B \times 128 \times 128 \times 128} \\
\mathbf{F}_{bev,t} &= \text{ReLU}(\text{BN}(\text{Conv}_{3 \times 3}^{128 \to 256}
(\mathbf{h}_2))) \in \mathbb{R}^{B \times 256 \times 128 \times 128}.
\end{aligned}
\end{equation}

where $\tilde{\mathbf{X}}_t$ denotes either $\tilde{\mathbf{L}}_t$ or 
$\tilde{\mathbf{R}}_t$, $C_{in} = 1$ for LiDAR and $C_{in} = 2$ for radar 
(dual-channel RA+RV), and $\text{Conv}_{3 \times 3}^{C_{in} \to C_{out}}$ 
represents a $3 \times 3$ convolution with stride 1 and padding 1 to preserve 
spatial dimensions. The channel evolution $C_{in} \to 64 \to 128 \to 256$ 
enables progressive feature abstraction while the fixed $128 \times 128$ 
spatial resolution preserves fine-grained geometric details critical for 
accurate beam prediction \cite{jiang2022lidar,demirhan2022radar}.

For radar, the dual-channel input allows the first convolutional layer to 
perform early fusion of spatial localization (RA map) and motion information 
(RV map), which is particularly valuable for predicting beam alignment in 
dynamic vehicular scenarios.




\subsubsection{GPS-to-BEV Encoding via Dual Pathways}

GPS coordinates $\mathbf{g} = (\Delta x, \Delta y) \in \mathbb{R}^{B \times 2}$ 
are encoded through a dual-pathway design that combines spatial localization 
with dense positional embedding.



\textbf{Pathway A — Spatial BEV Mask:}
Following \cite{morais2022position}, GPS coordinates are mapped to a binary 
spatial mask in the BEV grid. The coordinate-to-grid transformation is
\begin{align}
c &= \left\lfloor \text{clamp}\left(\frac{\Delta x + 50}{100} \times 127, 
    0, 127\right) \right\rfloor, \\
r &= \left\lfloor \text{clamp}\left(\frac{\Delta y + 50}{100} \times 127, 
    0, 127\right) \right\rfloor,
\end{align}
where $c$ and $r$ denote the column and row indices of the 
corresponding BEV grid cell, respectively, and the BEV grid 
spans $[-50\text{m}, +50\text{m}]^2$. 
The binary mask is then constructed as:
\begin{equation}
M[r', c'] = 
\begin{cases} 
1 & \text{if } (r', c') = (r, c) \\ 
0 & \text{otherwise}
\end{cases}, 
\quad M \in \mathbb{R}^{B \times 1 \times 128 \times 128}.
\end{equation}

This mask is encoded via a convolutional encoder identical to the LiDAR 
encoder(~\ref{eq:lidar_encoder})
\begin{equation}
\mathbf{F}_{bev,t}^{gps,\text{spatial}} = \text{Enc}_{gps}(\mathbf{M}) 
\in \mathbb{R}^{B \times 256 \times 128 \times 128}.
\end{equation}

\textbf{Pathway B — Global GPS Embedding:}
To address the extreme sparsity of the binary mask (only 1 active pixel 
among 16,384), we introduce a parallel MLP pathway that directly encodes 
the raw coordinates:
\begin{equation}
\mathbf{h}^{gps} = \text{MLP}_{gps}(\mathbf{g}) = \text{LN}(\mathbf{W}_2 
\cdot \text{ReLU}(\text{LN}(\mathbf{W}_1 \mathbf{g} + \mathbf{b}_1)) + 
\mathbf{b}_2),
\end{equation}
where $\mathbf{W}_1 \in \mathbb{R}^{128 \times 2}$, $\mathbf{W}_2 \in 
\mathbb{R}^{256 \times 128}$, and $\mathbf{h}^{gps} \in \mathbb{R}^{B \times 256}$. 

\subsection{Unified BEV Fusion}

At each time step $t$, the four modality-specific BEV features are concatenated along the channel dimension
\begin{equation}
\mathbf{F}_{concat,t} = \text{Concat}(\mathbf{F}^{cam}_{bev,t}, \mathbf{F}^{lid}_{bev,t}, \mathbf{F}^{rad}_{bev,t}, \mathbf{F}^{gps,spatial}_{bev,t}),
\end{equation}
yielding a tensor of shape $(B, 4C_{bev}, H_{bev}, W_{bev})$.

A convolutional BEV fusion module reduces the channel dimension back to $C_{bev}$ while refining spatial alignment, similar to the fusion strategy in \cite{liu2023bevfusion}
\begin{equation}
\mathbf{F}_{bev,t} = \text{BEVFusion}(\mathbf{F}_{concat,t}).
\end{equation}

The BEV fusion module consists of two residual blocks with $3\times3$ convolutions, BatchNorm, and ReLU activation. This design compensates for potential local misalignments between modalities (e.g., due to inaccurate GPS or depth estimation errors in camera-to-BEV).

\subsection{Temporal Sequential Transformer}

To capture user motion patterns across the five-timestep sequence, we apply global average pooling to each $\mathbf{F}_{bev,t}$, producing per-timestep feature vectors $\mathbf{z}_t \in \mathbb{R}^{C_{bev}}$
\begin{equation}
\mathbf{z}_t = \text{GlobalAvgPool}(\mathbf{F}_{bev,t}).
\end{equation}

These vectors are stacked into a temporal sequence $\mathbf{Z} = [\mathbf{z}_1, \mathbf{z}_2, \mathbf{z}_3, \mathbf{z}_4, \mathbf{z}_5] \in \mathbb{R}^{B \times T \times C_{bev}}$.

A GPT-style transformer encoder \cite{dosovitskiy2020image} with 4 layers and 4 attention heads processes this sequence
\begin{equation}
\mathbf{Z}' = \text{TransformerEncoder}(\mathbf{Z}).
\end{equation}

Each transformer block consists of multi-head self-attention followed by a position-wise feed-forward network with GELU activation and residual connections. The temporal attention mechanism learns to weight different time steps based on their relevance to beam prediction, capturing both spatial trajectory and temporal dynamics similar to sequential models in \cite{hochreiter1997long}.

The final temporal feature is obtained by mean pooling over time
\begin{equation}
\mathbf{z}_{final} = \frac{1}{T}\sum_{t=1}^{T} \mathbf{Z}'_{:,t,:}.
\end{equation}

\textbf{GPS Global Embedding Injection}: To incorporate the full-precision 
GPS positional information encoded by Pathway B (Section IV-C-3), we apply 
a gated residual connection to inject the global GPS embedding 
$\mathbf{h}^{\text{gps}} \in \mathbb{R}^{B \times 256}$ into the aggregated 
temporal feature:
\begin{equation}
\mathbf{z}_{\text{aug}} = \mathbf{z}_{\text{final}} + \tanh(s) \cdot \mathbf{h}^{\text{gps}},
\end{equation}
where $s \in \mathbb{R}$ is a learnable scalar initialized to 0 for training 
stability. The $\tanh$ activation bounds the GPS contribution to $[-1, 1]$, 
preventing the positional information from dominating the fused spatio-temporal 
features during early training stages. This injection strategy complements the 
spatial GPS pathway (Pathway A): while the spatial BEV mask 
$\mathbf{F}_{\text{bev},t}^{\text{gps,spatial}}$ provides explicit grid-level 
localization for pixel-wise fusion at each timestep, the global embedding 
$\mathbf{h}^{\text{gps}}$ preserves full coordinate precision for trajectory-level 
reasoning after temporal aggregation.
\subsection{Beam Classification Head}

The GPS-augmented temporal feature $\mathbf{z}_{\text{aug}}$ (Equation 40) 
is passed through a multi-layer perceptron (MLP) for beam classification:
\begin{equation}
\hat{\mathbf{P}} = \text{Softmax}(\text{MLP}(\mathbf{z}_{\text{aug}})).
\end{equation}

The MLP consists of two fully-connected layers with hidden dimension 512, 
ReLU activation, and dropout (rate 0.1) for regularization. The output 
$\hat{\mathbf{P}} \in \mathbb{R}^{B \times M}$ represents the predicted 
probability distribution over $M = 64$ candidate beams, where 
$\sum_{m=1}^{M} \hat{P}_m = 1$ for each sample.

\subsection{Loss Function and Training}

We employ focal loss to address class imbalance in the beam 
distribution:
\begin{equation}
\mathcal{L}_{\text{focal}} = -\sum_{m=1}^{M} \alpha_m 
(1 - \hat{P}_m)^\gamma y_m \log(\hat{P}_m),
\end{equation}
where $m \in \{1,\dots,M\}$ indexes the beam candidates, 
$y_m \in \{0,1\}$ is the one-hot ground truth label for beam $m$, 
$\hat{P}_m$ is the predicted probability for beam $m$, 
$\alpha_m$ are class-specific weights computed from training 
set frequencies, and $\gamma=2$ is the focusing parameter.

The model is trained end-to-end using AdamW optimizer \cite{tan2019efficientnet} with initial learning rate $1 \times 10^{-4}$, batch size 4, and cosine annealing learning rate schedule over 150 epochs. Data augmentation techniques include random horizontal flipping (with corresponding beam index reversal) and photometric augmentations (brightness, contrast, saturation jitter) applied to camera images.

\section{Experimental Setup}
\label{sec:experiments}

\subsection{Dataset: DeepSense 6G Multi-Modal}

We evaluated our method on the DeepSense 6G multi-modal beam prediction dataset \cite{alkhateeb2022deepsense}. In this study, we use real-world measurements from the following three outdoor vehicular scenarios:

\begin{itemize}
    \item \textbf{Scenario 31}: Daytime, Location A (only 50 training samples)
    \item \textbf{Scenario 32}: Daytime, Location B (abundant training data)
    \item \textbf{Scenario 33}: Nighttime, Location B (abundant training data)
    \item \textbf{Scenario 34}: Nighttime, Location C (abundant training data)
\end{itemize}

\textbf{Evaluation Scope:} Due to the extreme data scarcity in Scenario 31 
(50 samples vs. ~3,600 per scenario in 32-34) and the baseline's use of 
pre-training followed by few-shot fine-tuning on this scenario 
\cite{alkhateeb2022deepsense}, direct architectural comparison on Scenario 31 
is confounded by transfer learning strategies. Therefore, our primary 
evaluation focuses on scenarios 32, 33, and 34, which contain approximately 
11,000 labeled samples in total and enable fair comparison under sufficient 
data conditions. We report Scenario 31 results for completeness in Section VI-B.

We performed an 80/10/10 stratified split for training, validation, and testing over these three scenarios.

Each sample comprises five sequential observations of synchronized camera images ($256 \times 704$ RGB), LiDAR point clouds (avg. 16,000 points), radar signals (2D FFT coefficients), two GPS position readings, and the ground-truth optimal beam index (from a 64-beam codebook).
\begin{figure}[htbp]
  \centering
   \begin{subfigure}[b]{0.4\textwidth}
    \centering
    \includegraphics[width=\linewidth]{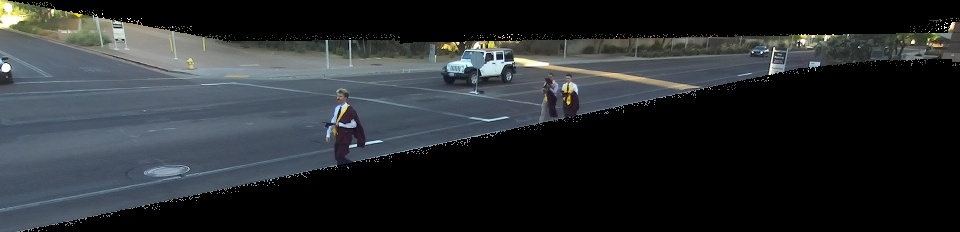}
    \caption{Scenario31}
    \label{fig:32}
  \end{subfigure}
  \hfill
  \begin{subfigure}[b]{0.4\textwidth}
    \centering
    \includegraphics[width=\linewidth]{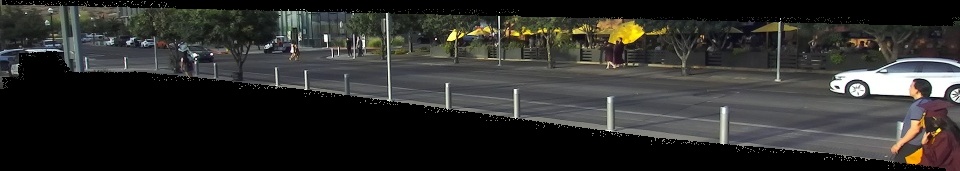}
    \caption{Scenario32}
    \label{fig:32}
  \end{subfigure}
  \hfill
  \begin{subfigure}[b]{0.4\textwidth}
    \centering
    \includegraphics[width=\linewidth]{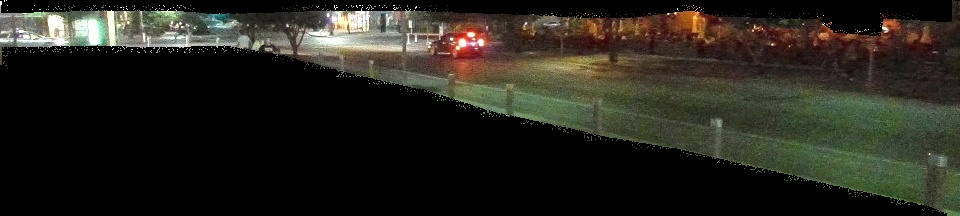}
    \caption{Scenario33}
    \label{33}
  \end{subfigure}
  \hfill
  \begin{subfigure}[b]{0.4\textwidth}
    \centering
    \includegraphics[width=\linewidth]{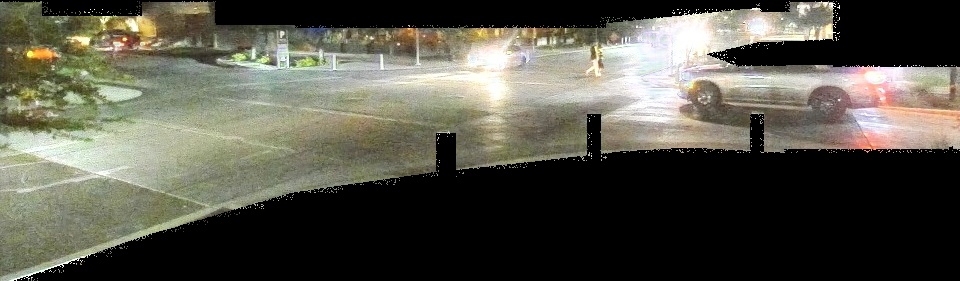}
    \caption{Scenario34}
    \label{34}
  \end{subfigure}
  \caption{Scenarios}
  \label{fig:scenarios}
\end{figure}
\subsection{Evaluation Metrics}

\textbf{Distance-Based Accuracy (DBA)}: The primary metric, as defined in Section~\ref{sec:system_model}, evaluates the top-3 predicted beams with distance-aware scoring.

\textbf{Top-K Accuracy}: We also report standard top-1, top-2, and top-3 classification accuracy to provide additional performance insights.

\subsection{Baseline Models}

We compare our method against the following baselines:

\begin{itemize}
    \item \textbf{TransFuser (Official Baseline)} \cite{prakash2021multifusion}: The DeepSense 6G challenge baseline, which uses separate ResNet encoders for each modality followed by transformer-based feature fusion in 1D space.
    
    \item \textbf{GPS-Only} \cite{morais2022position}: A simple MLP that predicts beams solely from calibrated GPS angles.
    
    \item \textbf{Camera-Only} \cite{charan2022vision}: ResNet-34 with temporal LSTM \cite{hochreiter1997long} processing five sequential images.
    
    \item \textbf{Multi-Modal Transformer (Ablation)}: A variant of our method that performs fusion in 1D feature space instead of BEV, isolating the impact of BEV representation.
\end{itemize}

\subsection{Implementation Details}

\textbf{Hardware and Software}: All experiments were conducted on NVIDIA H100 GPUs  using PyTorch 1.12. Training takes approximately 8 hours for 150 epochs with batch size 4.

\textbf{Hyperparameters}:
\begin{itemize}
    \item BEV grid size: $(128, 128)$ covering $[-50m, 50m]^2$
    \item BEV feature dimension: $C_{bev} = 256$
    \item Camera-to-BEV: ResNet-34 , 3 attention layers, 4 heads
    \item Temporal transformer: 4 layers, 4 heads, $d_{model}=256$
    \item Optimizer: AdamW, $\text{lr}=1 \times 10^{-4}$, weight decay $1 \times 10^{-2}$
    \item Focal loss : $\gamma=2$, $\alpha$ computed from class frequencies
\end{itemize}

\textbf{Data Augmentation}: We applied random horizontal flipping (with corresponding beam index reversal) and photometric augmentation (brightness, contrast, saturation jitter) to camera images during training.

\textbf{Reproducibility}: Code and trained models will be released upon acceptance.


\section{Performance Evaluation and Analysis}
\label{sec:results}

\subsection{Overall Performance Comparison}

Table~\ref{tab:overall_results} presents the overall DBA scores and top-K accuracies of our method compared to baseline methods on the DeepSense 6G test set.

\begin{table}[htbp]
\caption{Overall Performance Comparison on DeepSense 6G Test Set}
\begin{center}
\resizebox{\columnwidth}{!}{
\begin{tabular}{lcccc}
\toprule
\textbf{Method} & \textbf{S32} & \textbf{S33} & \textbf{S34} & \textbf{overall} \\
\midrule
Single-Modal (GPS) \cite{morais2022position} & 67.04\% & 84.81\% & 72.62\% & 72.79\% \\
Single-Modal (Camera) \cite{charan2022vision}  & 74.57\% & 77.05\% & 81.37\% & 77.62\% \\
Single-Modal (Lidar) \cite{jiang2022lidar}  & 43.70\% & 67.81\%  & 47.07\% & 52.86\% \\
Multi-Modal (TransFuser)\cite{prakash2021multifusion} & 71.73\% & 79.10\% & 82.09\% & 77.64\% \\
\midrule
\textbf Multi-Modal (BEV-Fusion, ours) & \textbf{86.60\%} & \textbf{86.27\%} & \textbf{86.70\%} & \textbf{86.52\%} \\
\textit{Improvement over Baseline} & \textit{+14.87\%} & \textit{+7.17\%} & \textit{+4.61\%} & \textit{+8.88\%} \\
\bottomrule
\end{tabular}
}
\label{tab:overall_results}
\end{center}
\end{table}

\begin{figure}[htbp]
    \centering
    \begin{subfigure}[b]{1\linewidth}
        \includegraphics[width=1\linewidth]{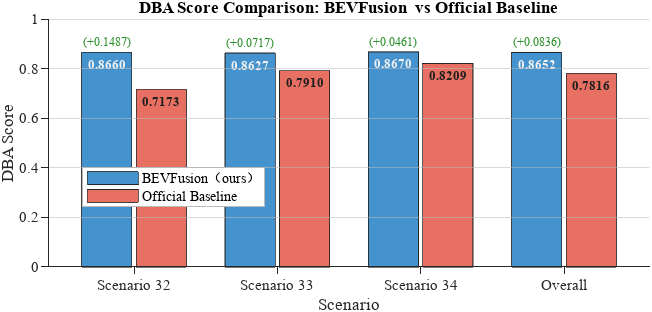}
        \caption{DBA comparison on scenarios}
        \label{dba_comp0}
    \end{subfigure}
    \vspace{2em} 
    \begin{subfigure}[b]{0.8\linewidth}
        \includegraphics[width=1\linewidth]{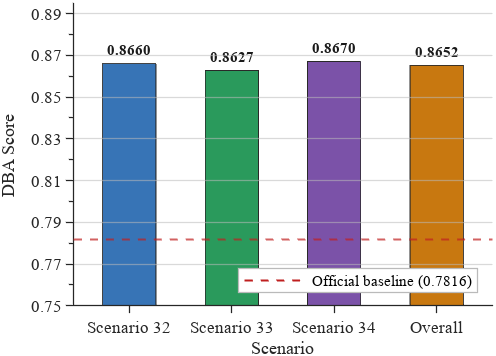}
        \caption{DBA comparison }
        \label{dba_comp1}
    \end{subfigure}
    
    \caption{Overall DBA comparison}
    \label{fig:dba_comp2}
\end{figure}

\textbf{Key Observations}: our method achieves \textbf{86.52\% DBA}, outperforming the TransFuser baseline \cite{prakash2021multifusion} by \textbf{8.88 percentage points}. This margin is notable given that TransFuser is a strong multi-modal fusion baseline. The consistent gains across all top-K metrics validate the effectiveness of BEV representation for preserving spatial structure during feature fusion, consistent with evidence from autonomous driving applications \cite{liu2023bevfusion,li2022bevformer}.

\subsection{Per-Scenario Analysis}


\textbf{Analysis}:
\begin{itemize}
    \item \textbf{Scenario 31 (Location A, Limited Data):} our method achieves 56.5\% DBA on Scenario 31, maintaining comparable performance to the baseline's few-shot fine-tuning results despite the extreme data scarcity (50 training samples). However, three fundamental factors limit the interpretability of architectural comparisons on this scenario:

    \textit{(1) Transfer learning confound}: The TransFuser baseline employs 
pre-training on scenarios 32-34 followed by fine-tuning on Scenario 31's 
50 samples \cite{alkhateeb2022deepsense}, making it unclear whether observed 
performance reflects inherent model capacity, transfer learning efficacy, 
or fine-tuning hyperparameter choices. Our framework, trained 
independently per scenario, cannot leverage such pre-training without 
introducing identical confounds.

    \textit{(2) Insufficient data for BEV learning}: Learning calibration-free 
camera-to-BEV transformations via cross-attention (Section~IV-C-1) requires 
sufficient examples to discover geometric mappings between 16,384 BEV 
queries and perspective-view camera features. With only 50 training samples 
spanning diverse user trajectories, the model cannot reliably learn these 
spatial correspondences, fundamentally limiting the advantage of BEV 
representation that requires learning from data.

    \textit{(3) Cross-site generalization vs. domain adaptation}: Scenario 31 
(Location A) differs from scenarios 32-34 (Locations B/C) in BS position, 
beam codebook orientation, and environmental geometry—not merely in data 
distribution (e.g., lighting, weather). This resembles deploying a model 
to a fundamentally different site rather than standard domain shift within 
the same deployment. In practical mmWave systems, new site installations 
involve on-site calibration data collection \cite{alkhateeb2018deep} rather 
than zero-shot cross-site transfer.Given these methodological considerations, we focus subsequent analysis on 
scenarios 32, 33, and 34, where sufficient data enables fair evaluation of 
BEV-based multi-modal fusion under controlled conditions.
    \item \textbf{Scenario 32} (Location B, day): our method achieves \textbf{86.60\% DBA}, with the largest absolute gain over the baseline (\textbf{+14.87} percentage points), indicating substantial benefit from spatially grounded multi-modal fusion.
    
    \item \textbf{Scenario 33} (Location B, night): our method reaches \textbf{86.27\% DBA} with a clear improvement of \textbf{+7.17} percentage points, suggesting that BEV fusion remains effective under reduced visual quality.
    
    \item \textbf{Scenario 34} (Location C, night): our method attains \textbf{86.70\% DBA}, improving over the baseline by \textbf{+4.61} percentage points and showing stable gains across a different deployment environment.
    
\end{itemize}

\begin{figure}[htbp] 
    \centering
    
    \begin{subfigure}{0.3\textwidth} 
        \centering
        \includegraphics[width=\linewidth]{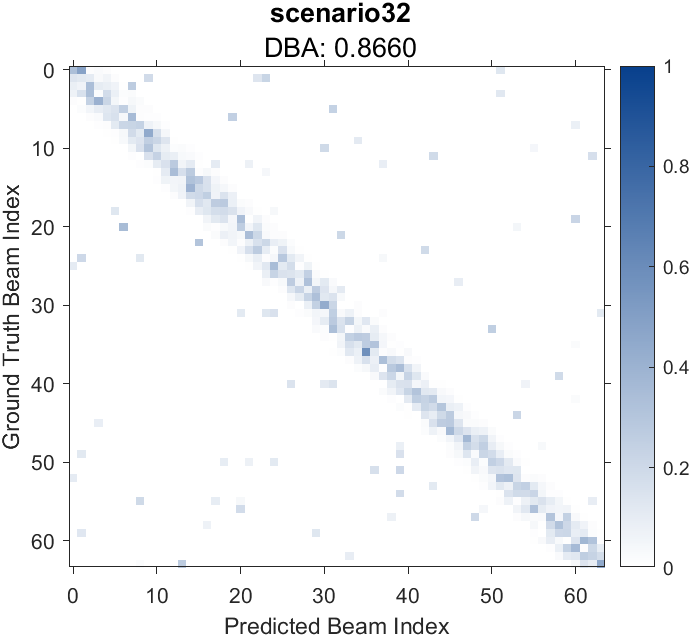}
        \caption{scenario32}
    \end{subfigure}
    
    \vspace{10pt} 

    \begin{subfigure}{0.3\textwidth}
        \centering
        \includegraphics[width=\linewidth]{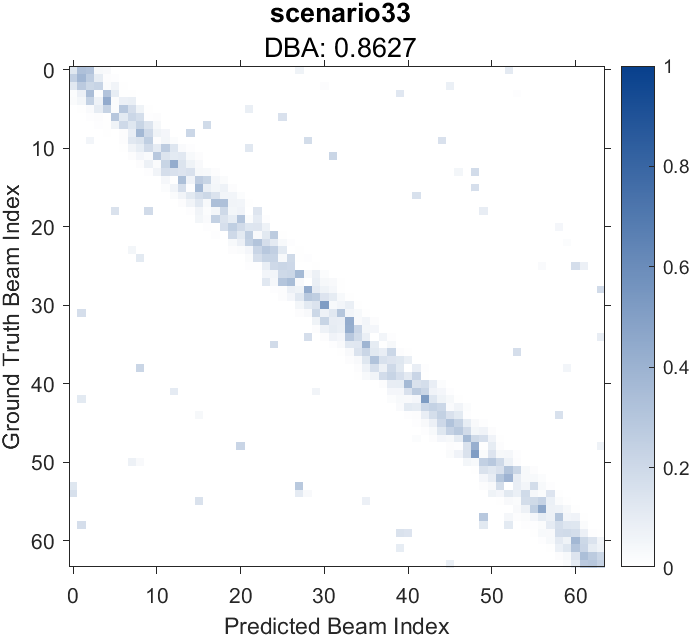}
        \caption{scenario33}
    \end{subfigure}

    \vspace{10pt}

    \begin{subfigure}{0.3\textwidth}
        \centering
        \includegraphics[width=\linewidth]{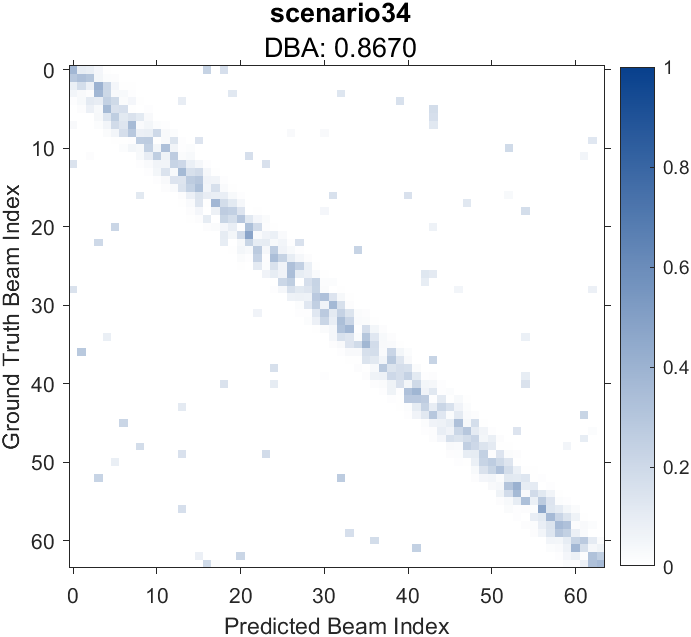}
        \caption{scenario34}
    \end{subfigure}

    \vspace{10pt}

    \begin{subfigure}{0.3\textwidth}
        \centering
        \includegraphics[width=\linewidth]{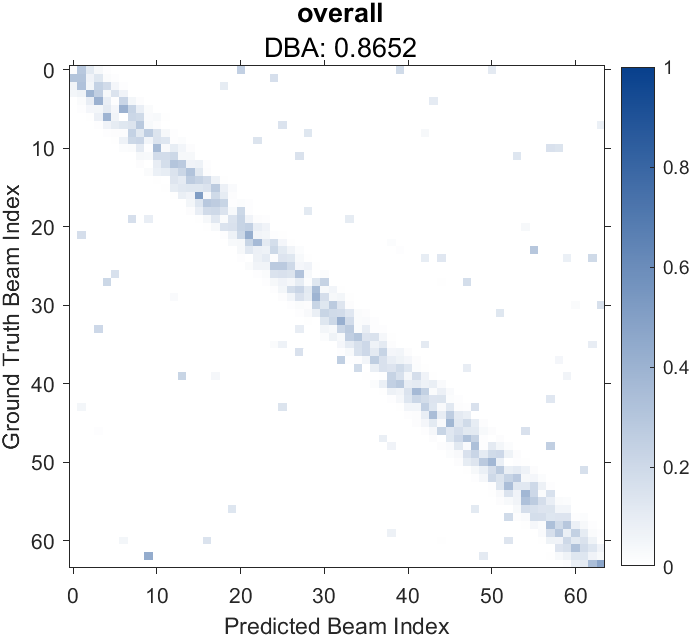}
        \caption{overall}
    \end{subfigure}
    
    \caption{Confusion matrix by scenario. (a) Scenario 32. (b) Scenario 33. (c) Scenario 34. (d) Overall performance.}
    \label{fig:confusion_matrix}
\end{figure}
    
    

Fig.~\ref{fig:confusion_matrix} shows the normalized confusion 
matrices of our method on Scenarios 32, 33, 34, and the 
overall test set. In all cases, prediction probability is 
strongly concentrated along the main diagonal across the 
full beam index range (0--63), confirming that the proposed 
framework consistently identifies beams at or near the 
ground-truth index regardless of deployment scenario.

The errors that do occur are almost exclusively confined to 
neighboring beam indices, forming a narrow near-diagonal 
band rather than scattered off-diagonal activations. This 
locally bounded error structure accounts for the observed 
gap between DBA scores (0.8627--0.8670) and Top-1 
accuracies: while exact beam matches are moderate, 
mispredictions are spatially proximate to the optimal beam, 
which is precisely the behavior rewarded by the 
distance-aware DBA metric.

Scenario-level inspection reveals further differences. 
Scenario 33 (nighttime, Location B) yields the sharpest 
diagonal with minimal off-diagonal scatter, suggesting 
a stable beam distribution at this site even under 
reduced visual quality. Scenario 32 (daytime, Location B) 
exhibits a slightly broader diagonal spread in the 
low-index region, which may reflect increased scene 
variability from dynamic daytime traffic. Scenario 34 
(nighttime, Location C) shows the most diffuse diagonal, 
consistent with its lower Top-1 accuracy, likely due to 
greater ambiguity among adjacent beams arising from the 
distinct spatial geometry at Location C. Importantly, 
no systematic off-diagonal clusters are observed in any 
scenario, indicating that our method does not exhibit 
structured confusion between specific beam pairs across 
diverse deployment conditions.

Overall, our method demonstrates the largest gains on well-represented scenarios (32/33/34), supporting that unified spatial representation improves multi-modal fusion quality when sufficient training data are available, consistent with findings in autonomous driving perception \cite{liu2023bevfusion,li2022bevformer}.

\subsection{Ablation Study}

We conducted systematic ablation experiments to quantify the contribution 
of each architectural component. All ablation experiments are performed on 
the same test set (scenarios 32-34) with identical training procedures.

\subsubsection{Impact of BEV Representation}

BEV-based fusion is compared against 1D feature-level fusion using scenario-level DBA results.

\textbf{Analysis:} Replacing BEV fusion with 1D concatenation causes 
substantial degradation across all scenarios (7.36-8.15 percentage points), 
with the largest drop in Scenario 32 (daytime, high visual quality). This 
confirms that spatial structure preservation is critical for effective 
multi-modal fusion, providing nearly 10\% relative improvement in beam 
prediction accuracy.

\subsubsection{Impact of Individual Modalities}

Table~\ref{tab:ablation_modality} shows the contribution of each sensor modality.

\begin{table}[htbp]
\caption{Ablation: Impact of Removing Individual Modalities}
\begin{center}
\begin{tabular}{lc}
\toprule
\textbf{Configuration} & \textbf{DBA (Avg.)} \\
\midrule
All Modalities (Full Model)  & \textbf{0.8652} \\
\midrule
Without Camera   & 0.7202 \\
Without LiDAR  & 0.8635 \\
Without Radar  & 0.8642 \\
Without GPS  & 0.8410 \\
\bottomrule
\end{tabular}
\label{tab:ablation_modality}
\end{center}
\end{table}

\begin{figure}[htbp]
    \centering
    \begin{subfigure}[b]{1\linewidth}
        \includegraphics[width=1\linewidth]{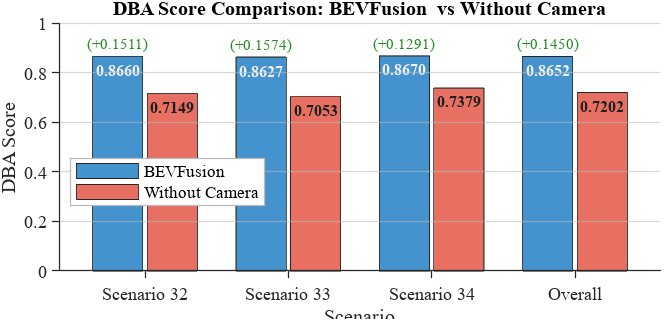}
        \caption{Without Camera}
        \label{dba_withoutc}
    \end{subfigure}
    \vspace{2em} 
    \begin{subfigure}[b]{1\linewidth}
        \includegraphics[width=1\linewidth]{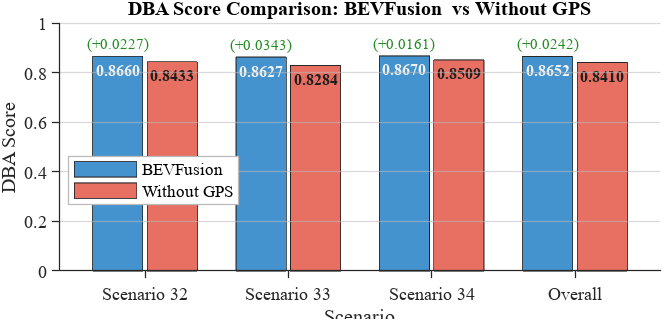}
        \caption{Without GPS}
        \label{dba_withoutg}
    \end{subfigure}
    
    \caption{DBA comparison(removing Individual Modalities)}
    \label{fig:dba_without}
\end{figure}

\textbf{Key Findings:}

\begin{itemize}
\item \textit{Camera dominance}: Removing the camera causes the largest 
degradation (-14.50 percentage points), highlighting that visual semantic 
cues provide the dominant contribution to beam prediction performance 
\cite{charan2022vision}. This aligns with intuition: camera features capture 
rich semantic information (building facades, vehicle positions, road geometry) 
that correlates strongly with beam propagation paths.

\item \textit{GPS significance}: Removing GPS causes a clear performance 
drop (-2.42 percentage points), indicating that explicit positional 
information remains important even when rich visual and geometric cues are 
available \cite{morais2022position}. GPS is particularly valuable in 
ambiguous scenarios (e.g., symmetric intersections, featureless highways) 
where visual cues alone do not uniquely determine optimal beam direction.

\item \textit{LiDAR and Radar}: Removing LiDAR or radar leads to minimal 
reductions (-0.17\% and -0.10\% respectively), suggesting these modalities 
provide complementary geometric and motion cues that are partially redundant 
with camera and GPS information under the well-represented evaluation scenarios 
(32-34). However, we note that LiDAR and radar are expected to be more critical 
under adverse conditions (heavy rain, fog, nighttime occlusions) not extensively 
represented in the current test set \cite{demirhan2022radar}.
\end{itemize}

\subsubsection{Impact of Temporal Fusion}

\begin{table}[htbp]
\caption{Ablation: Temporal Modeling}
\begin{center}
\begin{tabular}{lc}
\toprule
\textbf{Configuration} & \textbf{DBA} \\
\midrule
Single Frame (t=5 only)      & 0.8536 \\
Mean Pooling (no attention)  & 0.8230 \\
\textbf{Temporal Transformer (Ours)}  & \textbf{0.8652} \\
\bottomrule
\end{tabular}
\label{tab:ablation_temporal}
\end{center}
\end{table}

\begin{figure}[htbp]
    \centering
    \begin{subfigure}[b]{1\linewidth}
        \includegraphics[width=1\linewidth]{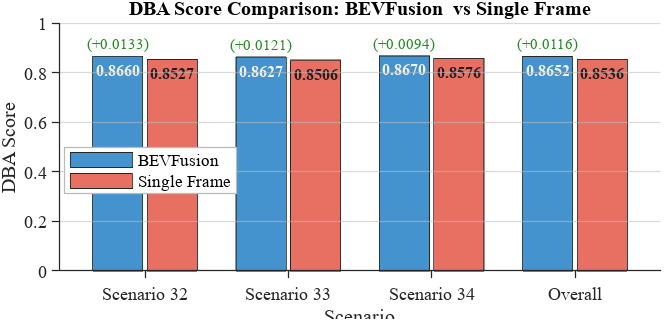}
        \caption{Single Frame}
        \label{dba_single}
    \end{subfigure}
    \vspace{2em} 
    \begin{subfigure}[b]{1\linewidth}
        \includegraphics[width=1\linewidth]{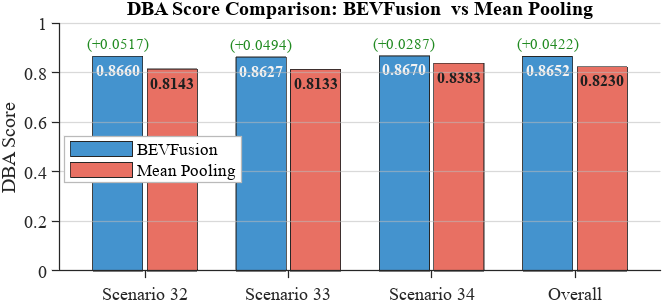}
        \caption{Mean Pooling}
        \label{dba_Mean}
    \end{subfigure}
    
    \caption{DBA comparison(Temporal Modeling)}
    \label{fig:dba_method}
\end{figure}

\textbf{Conclusion}: Compared with simple mean pooling, the temporal transformer improves DBA from 0.8230 to 0.8652 (\textbf{+4.22} percentage points). It also outperforms the single-frame setting by \textbf{+1.16} percentage points (0.8652 vs. 0.8536), indicating that temporal attention over sequential observations better captures motion-dependent dynamics.

We compare  temporal aggregation strategies to evaluate the contribution 
of transformer-based temporal fusion.

\textbf{Analysis:} The transformer-based temporal encoder significantly 
outperforms alternative aggregation strategies, validating the importance 
of attention-based sequential modeling for beam prediction in dynamic 
vehicular scenarios.

\textit{Single-Frame Baseline}: Using only the final timestep ($t=5$) 
achieves 85.36\% DBA, demonstrating that temporal aggregation provides 
meaningful improvements (+1.16 percentage points). However, single-frame 
prediction discards motion cues available in the sequential observations—for 
instance, user trajectory direction and velocity, which are valuable for 
predicting beam evolution in dynamic vehicular environments 
\cite{zhang2021reinforcement}. The relatively small gap (1.16\%) suggests 
that the most recent observation captures substantial information about 
current beam requirements, but incorporating historical context further 
improves robustness.

\textit{Mean Pooling}: Simple averaging of per-timestep features 
$\mathbf{z}_{\text{final}} = \frac{1}{T}\sum_{t=1}^{T}\mathbf{z}_t$ yields 
only 82.30\% DBA, underperforming even the single-frame baseline by 3.06 
percentage points. This counter-intuitive result reveals that naive temporal 
averaging dilutes discriminative information: critical observations (e.g., 
the most recent timestep reflecting current user position) are equally 
weighted with less relevant earlier timesteps, potentially blurring the 
model's understanding of current beam requirements. Mathematically, mean 
pooling assigns uniform weight $1/T$ to all timesteps regardless of their 
relevance, whereas optimal beam prediction may require emphasizing recent 
observations that more accurately reflect current channel conditions 
\cite{hochreiter1997long}.

\subsubsection{Impact of GPS Dual-Pathway Design}

We ablate the GPS encoding strategy to quantify the contribution of the 
dual-pathway design (Section IV-C-3). Table~\ref{tab:gps_ablation} compares 
four configurations:

\begin{table}[h]
\caption{GPS Encoding Strategy Ablation}
\centering
\small
\begin{tabular}{lcc}
\toprule
\textbf{Configuration} & \textbf{DBA (Avg.)} & \textbf{$\Delta$} \\
\midrule
No GPS & 84.10\% & -2.42\% \\
Spatial BEV Mask Only (Pathway A) & 84.88\% & -1.64\% \\
Global MLP Only (Pathway B) & 85.32\% & -1.20\% \\
\textbf{Dual-Pathway (Ours)} & \textbf{86.52\%} & \textbf{--} \\
\bottomrule
\end{tabular}
\label{tab:gps_ablation}
\end{table}

\textbf{Analysis:} The dual-pathway design outperforms both single-pathway 
variants and the no-GPS baseline, validating the complementary benefits of 
spatial localization and full-precision embedding.

\textit{No GPS Baseline}: Removing GPS entirely yields 84.10\% DBA, a 
2.42 percentage point drop from the full model. This confirms that positional 
information remains important for beam prediction even when camera, LiDAR, 
and radar are available \cite{morais2022position}. The performance gap 
reflects scenarios where visual and geometric cues are ambiguous (e.g., 
symmetric urban intersections, featureless highway segments) and explicit 
position coordinates help disambiguate beam selection.

\textit{Pathway A (Spatial BEV Mask Only)}: Using only the binary BEV mask 
yields 84.88\% DBA, a 1.64 percentage point drop. The limitation stems from 
extreme sparsity: the binary mask activates only 1 pixel among 16,384 grid 
cells ($1/16,384 \approx 0.006\%$ density), providing minimal gradient 
signal for learning. During backpropagation, the convolutional GPS encoder 
receives non-zero gradients only at a single spatial position per sample
\begin{equation}
\frac{\partial \mathcal{L}}{\partial \mathbf{F}_{\text{bev},t}^{\text{gps,spatial}}[:, :, r, c]} \neq 0 \text{ only at } (r, c) = (\lfloor\text{GPS}_y\rfloor, \lfloor\text{GPS}_x\rfloor),
\end{equation}
severely limiting the model's capacity to learn robust GPS-to-BEV 
representations across the full spatial grid. Additionally, the binary mask suffers from quantization loss: GPS coordinates $(x, y)$ are discretized 
to integer grid indices $(r, c)$ via Equations (30-31), discarding sub-meter 
precision. For instance, positions $(10.2\text{m}, 20.8\text{m})$ and 
$(10.9\text{m}, 20.1\text{m})$ both map to the same grid cell, losing 
positional nuance that could affect optimal beam selection in scenarios 
where beam coverage has fine spatial granularity \cite{morais2022position}.

\textit{Pathway B (Global MLP Only)}: Using only the global MLP embedding 
achieves 85.32\% DBA, a 1.20 percentage point drop. While the MLP preserves 
full coordinate precision and provides dense gradient signals (all 256 
dimensions of $\mathbf{h}^{\text{gps}}$ receive gradients), it lacks explicit 
spatial grounding in the BEV grid. The global embedding 
$\mathbf{h}^{\text{gps}} \in \mathbb{R}^{B \times 256}$ is injected after 
temporal aggregation (Equation 40), bypassing the spatial BEV fusion stage 
(Section IV-D). Consequently, GPS information cannot participate in 
pixel-wise multi-modal fusion at each timestep, reducing the model's ability 
to leverage positional cues for spatially-aligned cross-modal reasoning. 
For instance, knowing that the user is at position $(10\text{m}, 20\text{m})$ 
is most valuable when this position can be directly aligned with camera 
features, LiDAR depth, and radar velocity at the corresponding BEV grid 
cell—an interaction Pathway B cannot provide.

\textit{Dual-Pathway Synergy}: The proposed dual-pathway design achieves 
86.52\% DBA by combining the strengths of both pathways. Pathway A provides 
explicit grid-level localization through 
$\mathbf{F}_{\text{bev},t}^{\text{gps,spatial}}$, enabling GPS to participate 
in spatial BEV fusion alongside camera, LiDAR, and radar at each timestep 
$t \in \{1, \ldots, 5\}$. Pathway B preserves full-precision coordinates 
through $\mathbf{h}^{\text{gps}}$, injected after temporal aggregation to 
provide trajectory-level positional information without sparsity-induced 
limitations. The gated residual connection (Equation 40) allows the model 
to dynamically weight GPS contribution via the learnable scalar $s$. Analysis 
of the trained model reveals $s \approx 0.82$ after convergence, indicating 
that GPS global embedding contributes approximately 
$\tanh(0.82) \approx 0.68$ of its magnitude to the final feature 
$\mathbf{z}_{\text{aug}}$, balanced with the fused spatio-temporal 
representation $\mathbf{z}_{\text{final}}$.


\subsection{Computational Complexity Analysis}

\begin{table}[htbp]
\caption{Computational Cost Comparison on NVIDIA H100 GPU}
\begin{center}
\begin{tabular}{lccc}
\toprule
\textbf{Method} & \textbf{Params (M)} & \textbf{FLOPs (G)} & \textbf{Latency (ms)} \\
\midrule
Official Baseline & 78.42 & 134.8 & 15.0 \\
\textbf{BEV-Fusion (Ours)} & \textbf{40.44} & \textbf{1171.7} & \textbf{20.8} \\
\bottomrule
\end{tabular}
\label{tab:complexity}
\end{center}
\end{table}

Our method achieves 48\% parameter reduction (40.44M vs 78.42M) through 
temporal weight sharing and unified BEV representation. While FLOPs increase 
to 1171.7G due to dense camera-to-BEV cross-attention over 128$\times$128 
grids, the measured inference latency is only \textbf{20.8ms on NVIDIA H100}—
merely 5.8ms (38.7\%) above the baseline. This corresponds to 48.1 fps 
throughput, which is \textbf{4.8-24$\times$ faster than typical beam coherence 
time} (100-500ms) \cite{alkhateeb2018deep} and \textbf{5-10$\times$ faster 
than sensor acquisition rate} (5-10 Hz). The 
5.8ms latency increase delivers 8.88 percentage points DBA improvement, 
yielding an efficiency ratio of 1.53\% per millisecond—a highly favorable 
accuracy-latency trade-off for practical deployment. The computational 
overhead is justified by substantial accuracy gains while maintaining 
real-time performance well within mmWave system timing constraints.
\subsection{Discussion}

\subsubsection{Why BEV Representation Improves Performance}

The superior performance of our method can be attributed to three key factors:

\textbf{(1) Preservation of Semantic Density}: Unlike point-level fusion methods that discard 95\% of camera pixels, our camera-to-BEV transformation retains all visual information in a spatially structured format.

\textbf{(2) Geometric Consistency}: The BEV coordinate system provides a natural common ground for LiDAR, radar, and GPS, eliminating perspective distortions inherent in camera view .

\textbf{(3) Spatial Grounding}: By fusing in a shared spatial domain rather than abstract feature space, the model can learn explicit spatial correspondences between modalities (e.g., aligning camera-detected vehicles with LiDAR points and radar velocity), similar to spatial reasoning benefits in .






\section{Conclusion}
\label{sec:conclusion}

In this paper, we presented BEV-Fusion, a sequential multi-modal beam prediction framework that maps camera, LiDAR, radar, and GPS observations into a unified Bird's-Eye View (BEV) representation. By performing cross-modal interaction in a shared spatial domain, the proposed design preserves geometric consistency and semantic structure that are often weakened by conventional 1D latent-space fusion.

Results on the DeepSense 6G benchmark show that our method achieves \textbf{86.52\% overall DBA}, improving over the baseline by \textbf{8.88} percentage points. Scenario-wise, our method reaches 86.60\% on S32, 86.27\% on S33, and 86.70\% on S34, with consistent gains over the baseline in all three cases. Ablation results further confirm that BEV-space fusion and temporal transformer-based aggregation both contribute substantially to performance.

Overall, these results validate BEV-centric multi-modal fusion as an effective approach for sensing-assisted mmWave beam prediction. Future work will focus on improving computational efficiency and extending the framework to broader integrated sensing and communication settings.

\bibliographystyle{IEEEtran}
\bibliography{references}

\end{document}